\documentclass[runningheads]{svmult}

\usepackage{makeidx}   % allows index generation
\usepackage{graphicx}  % standard LaTeX graphics tool
\usepackage{subeqnar}  % subnumbers individual equations
\usepackage{multicol}  % used for the two-column index
\usepackage{physprbb}  % modified textarea for proceedings,
\makeindex             % used for the subject index
\begin{document}
\title*{Thermonuclear Supernovae: Is Deflagration Triggered by Floating Bubbles?}
\toctitle{Thermonuclear Supernovae: Is Deflagration Triggered by Floating 
Bubbles?}
\titlerunning{Is Deflagration Triggered by Floating Bubbles?}
\author{Eduardo Bravo\inst{1,2}
\and Domingo Garc\'\i a-Senz\inst{1,2}
}
\authorrunning{Eduardo Bravo \& Domingo Garc\'\i a-Senz}
\institute{
	Dpt. F\'\i sica i Eng. Nuclear, 
	Universitat Polit\`ecnica de Catalunya,
	Diagonal 647, \\
	08028 Barcelona, Spain
\and 
	Institut d'Estudis Espacials de Catalunya,
	Gran Capit\`a 2-4, 
	08034 Barcelona
}

\maketitle              % typesets the title of the contribution

\begin{abstract}
In recent years, it has become clear from multidimensional 
simulations that the outcome of deflagrations depends strongly on the initial 
configuration of the flame. 
We have studied under which conditions this configuration could consist of a 
number of scattered, isolated, hot bubbles. Afterwards, we have calculated the 
evolution of deflagrations starting from different numbers of bubbles. We have 
found that
starting from 30 bubbles a mild explosion is produced
($M (^{56}{\mathrm Ni}) = 0.56 {\mathrm M}_\odot$), while starting from 10 
bubbles the star becomes only marginally unbound ($K = 0.05$ foes).
\end{abstract}

\section{The initial configuration of deflagrations in white dwarfs and its
outcomes}
\index{thermonuclear supernovae}
\index{type Ia supernovae}
\index{delayed detonation}
\index{deflagration}
\index{supernovae: initial conditions}
\index{bubbles}
\index{multidimensional hydrodynamics}
The explosion mechanism(s) responsible for thermonuclear supernovae (SNIa) is
still not well known. Although there have been recently some claims that the
delayed detonation mechanism lacks a physical background, there are still
unexplored mechanisms by which the transition from deflagration to detonation could occur (see Garc\'\i a-Senz \& Bravo, this volume). Here we concentrate on the
other possibility, i.e. a pure deflagration that would process about a solar mass,
synthesizing of the order of $0.5 {\mathrm M}_\odot$ of $^{56}$Ni in
order to make a typical SNIa. 
Multidimensional calculations of deflagrations 
bear the advantage over one-dimensional models that the energy generation rate
becomes eventually independent from the local value of the
flame speed (\cite{rhn02,k00}, see also Niemeyer, this volume). In these calculations, the 
acceleration of the fuel consumption rate is due to the deformation 
of the flame surface, which is well accounted for by 3D hydrocodes. 

In this picture, the main free
parameter is the initial configuration
of the flame. There have been a few \cite{hs02} simulations of
the transition from the hydrostatic phase to the
hydrodynamic one, but a self consistent multidimensional initial model is still
far from being available. However, a plausible ignition scenario was suggested in
\cite{gw95}: the nearly-simultaneous runaway at several different
spots (from here on, bubbles) in the central region of the white dwarf. In this 
paper, we explore the dependence of the outcome of the explosion on
the initial number of bubbles. First, we address the statistics of the initial 
distribution of bubble radii and, then, we present the results of a couple of
hydrodynamical simulations performed with an SPH code (\cite{gbs98} and Garc\'\i a-Senz \& Bravo, this volume). 

\subsection{Statistical approach to the initial distribution of igniting 
bubbles}
\index{ignition timescale}
Our statistical approach is based on the following assumptions:
{\bf A)} As a result of convection, there appear a number of hot spots, characterized by its central 
(peak) temperature, $T_0$, and its
thermal profile (which we take here exponential, with characteristic exponent 
$R_0$). 
{\bf B)} Each hot spot evolves adiabatically in place, in a time given by its ignition timescale. 
{\bf C)} The peak temperatures of the hot spots can be characterized by a
statistical continuous function, $\theta\left(T_0\right)$.

Depending on the thermal gradient inside each hot spot our results can be
split into two different regimes:
{\bf 1)} If the thermal profile is shallow enough, bubbles grow
due to spontaneous flame propagation, {\bf 2)} otherwise, bubbles
grow conductively. In the first case, the distribution function of bubble radii
is a time function given by
\begin{equation}
{{\D N}\over{\D R_{\mathrm b}}} = \theta\left(T_0\right) {{A^{1/B}\cdot10^9}
\over{R_0}}
{{\tau_{\mathrm i}\left(T_{\mathrm 0b}\right) \exp\left(B R_{\mathrm
b}/R_0\right)}\over{\left\{t+\tau_{\mathrm i}\left(T_{\mathrm 0b}\right) 
\left[1-\exp\left(B R_{\mathrm b}/R_0\right)\right]\right\}^{1+1/B}}}\; ,
\end{equation}
where $\tau_{\mathrm i}\left(T_{\mathrm 0b}\right)\sim A \left(T_{\mathrm
0b}/10^9~{\mathrm K}\right)^{-B}$ is
the ignition timescale, and $T_{\mathrm 0b} = 10^9$~K. In the second case, the
distribution function becomes
\begin{equation}
{{\D N}\over{\D R_{\mathrm b}}} = \theta\left(T_0\right) {{A^{1/B}}\over{B
v_{\mathrm cond}}} \left(t-{{R_{\mathrm b}}\over{v_{\mathrm
cond}}}\right)^{1+1/B}\; .
\end{equation}
In the first case, and for any reasonable choice of the function $\theta$, the resulting distribution has a sharp peak at a determined
value of $R_{\mathrm b}$ so that the initial
configuration is composed of an arbitrary number of {\it equal size bubbles}. 
In contrast, in the second case the bubble radii distribution is continuous up 
to a maximum radius, which results in a initial
distribution of {\it unequal size bubbles}. 

\subsection{SPH simulations of deflagrations triggered by floating bubbles}
\begin{table}[b]
\caption{Model parameters and results}
\begin{center}
\renewcommand{\arraystretch}{1.4}
\setlength\tabcolsep{5pt}
\begin{tabular}{llllllll}
\hline\noalign{\smallskip}
Model & $N_{\mathrm b}$ & $R_{\mathrm b}$ & $M_{\mathrm b}$ & 
$E_{\mathrm kin}$ & $M_{\mathrm Ni}$ & $M_{\mathrm C+O}$ & $M_{\mathrm Si}$ \\
 & & $\left(10^6 {\mathrm cm}\right)$ & & $\left(10^{51} {\mathrm erg}\right)$ &
 $\left({\mathrm M}_\odot \right)$ & $\left({\mathrm M}_\odot \right)$ & 
 $\left({\mathrm M}_\odot \right)$ \\
\noalign{\smallskip}
\hline
\noalign{\smallskip}
B30U & 30 & 6.7 & 2.8\% & 0.44 & 0.56 & 0.65 & 0.05 \\
B10U & 10 & 9.7 & 3\% & 0.05 & 0.24 & 0.88 & 0.04 \\
\hline
\end{tabular}
\end{center}
\label{tab1}
\end{table}

\begin{figure}[t]
\begin{center}
\includegraphics[width=.85\textwidth]{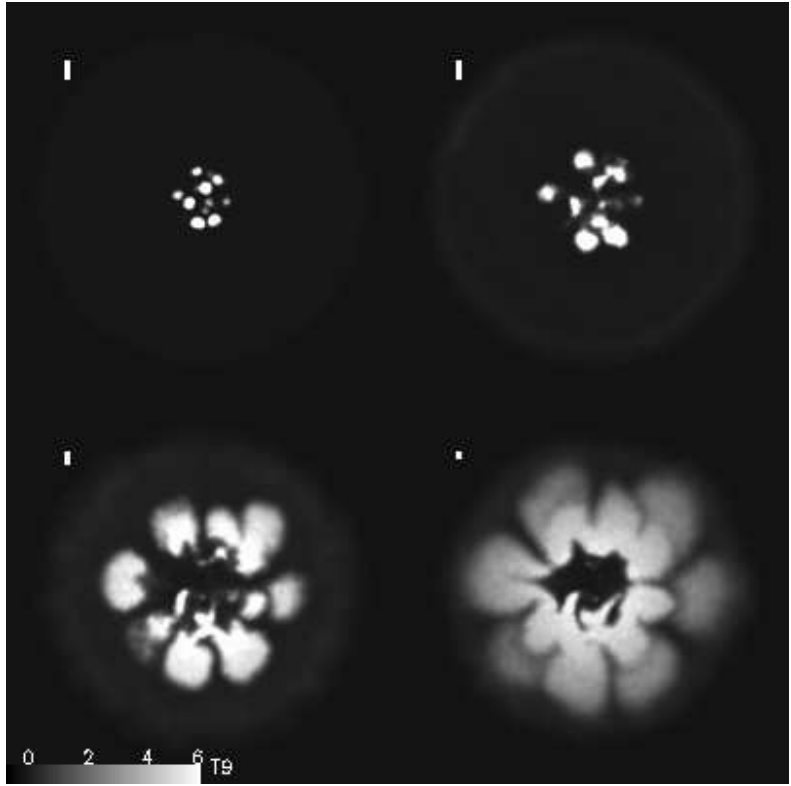}
\end{center}
\caption{Snapshots of the temperature distribution in a meridian plane at
times 0, 0.27, 0.58, and 0.81~s for model B30U. The temperature scale is shown 
at the bottom
left of the image, while the length scale is shown at the top left of each 
snapshot (the
length of the vertical bar is equivalent to 200~km)}
\label{fig1}
\end{figure}

\begin{figure}[bt]
\begin{center}
{
 \centering 
 \leavevmode 
 \columnwidth=.45\columnwidth 
 \includegraphics[width=\columnwidth]{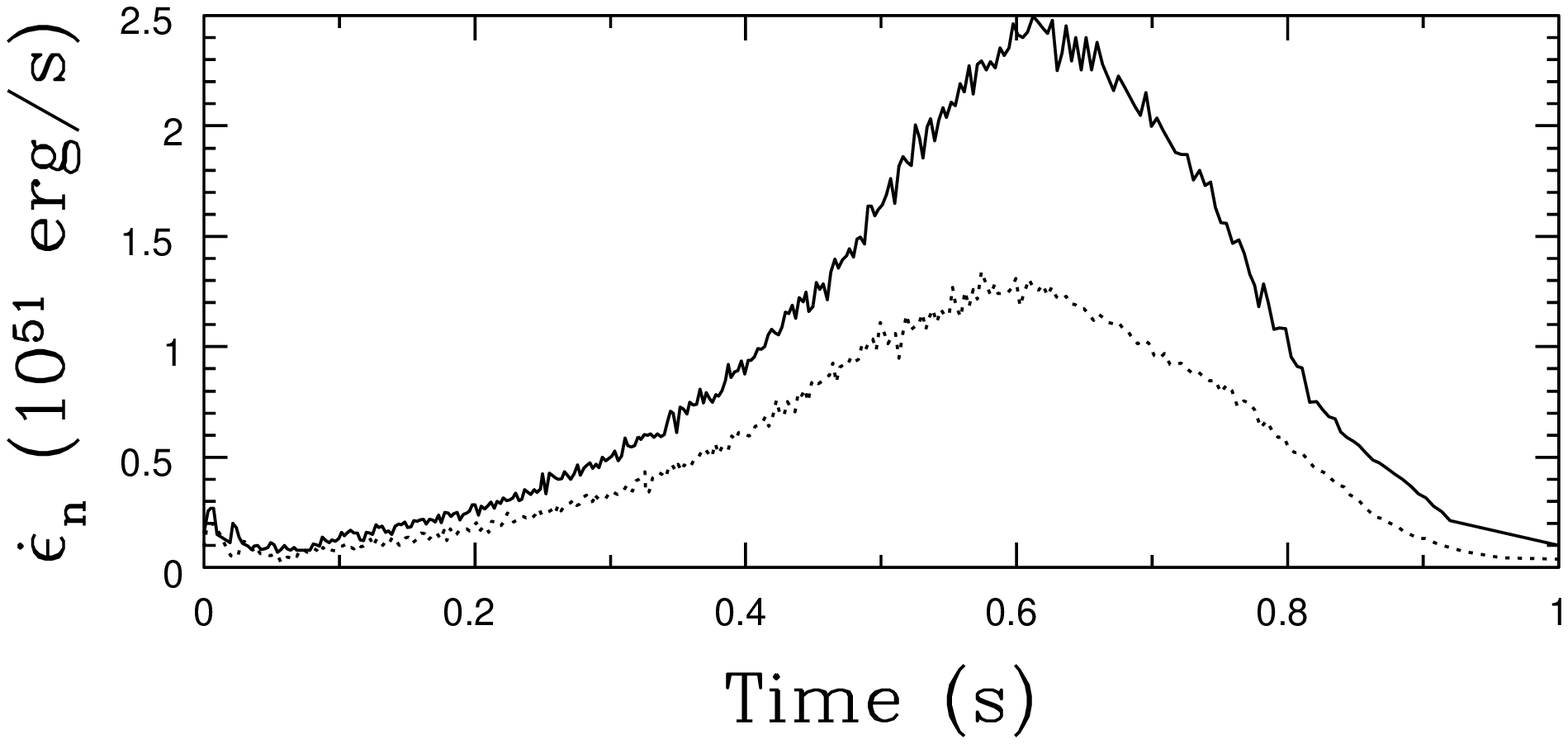}
 \hfil 
 \includegraphics[width=\columnwidth]{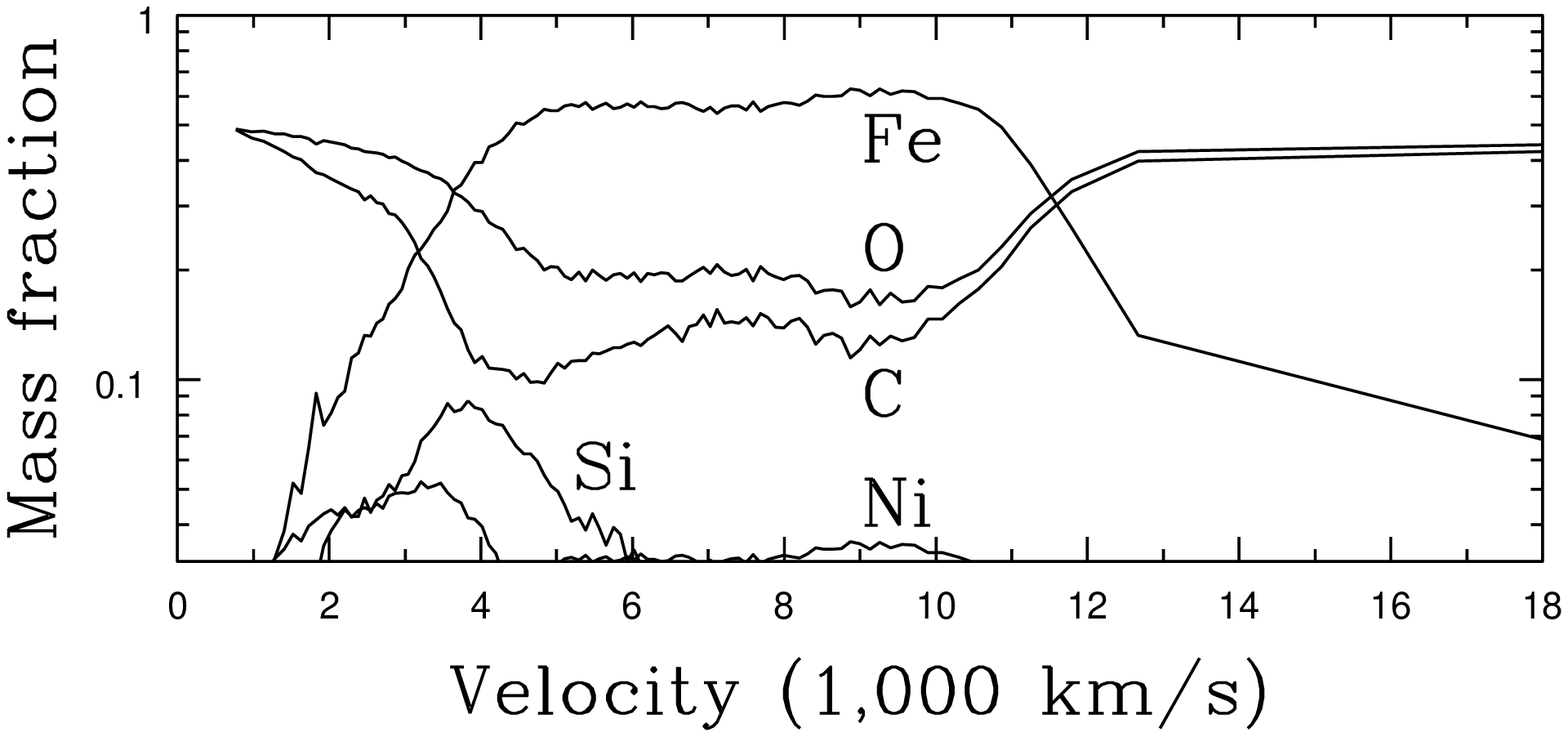}
}
\end{center}
\caption{Results of the hydrodynamical calculation.
(\textbf{a}) Nuclear energy generation rate as a function of time for models
B30U ({\it solid line}) and B10U ({\it dotted line}). 
(\textbf{b}) Final distribution of elements in velocity space for model B30U
}
\label{fig2}
\end{figure}

\begin{figure}[tb]
\begin{center}
\includegraphics[width=.3\textwidth, angle=270]{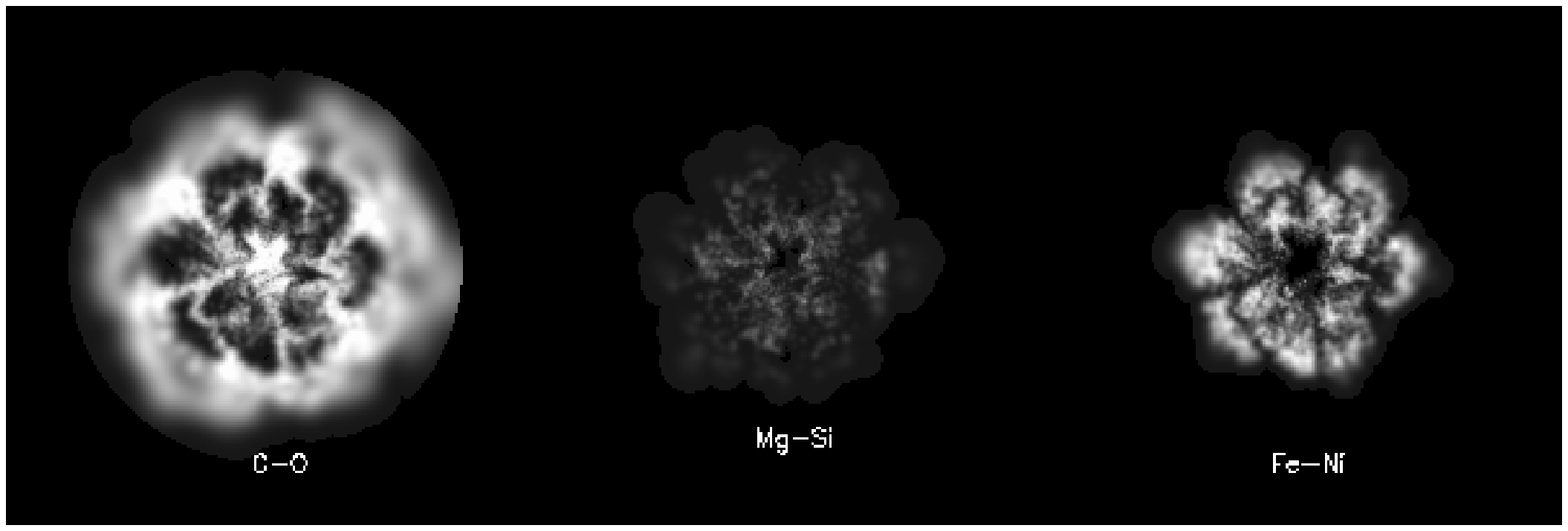}
\end{center}
\caption{Spatial distribution of the main chemical species at the end of the
calculation of model B30U, in the same meridian plane as in Fig.~\ref{fig1}}
\label{fig3}
\end{figure}

Given the constraints that the value of $R_0$ must satisfy in order to get an
initial configuration composed of identical bubbles, we estimate as quite
improbable this case. However, for simplicity, here we have adopted this
configuration as the starting point of our simulations. We have followed the
evolution of the explosion of a Chandrasekhar mass white dwarf of initial
density $1.9\,10^9$~g/cm$^3$ starting from two different numbers of equal size
bubbles, as detailed in Table~\ref{tab1} ($N_{\mathrm b}$ is the initial
number of bubbles, $R_{\mathrm b}$ its initial radius, $M_{\mathrm b}$ the 
mass incinerated initially, and the other symbols have their usual meanings). The
calculation was performed in 3D with the above mentioned SPH code using 250,000
particles, with a central resolution of $20$~km. It is important to emphasize 
that we did not impose any
artificial symmetry conditions and, thus, our model has no artificial 
characteristic \index{symmetry conditions}
length other than the own resolution of the code. Actually, the imposition of
artificial symmetry conditions (for instance, symmetry planes) in any 3D 
hydrodynamic
calculation biases the development of large-scale structures or even of the
small-scale ones if they are close enough to a symmetry plane.

The results of our calculations can be found in Table~\ref{tab1} and in 
Figs.~\ref{fig1} and \ref{fig2}a. The outcome depends strongly on the initial
number of bubbles. The deflagration is powered by the evolution and 
interaction of the bubbles:
growth, buoyancy (second snapshot in Fig.~\ref{fig1}), and merging (third
snapshot). The maximum acceleration of combustion is obtained when the bubbles
interact with each other, feeding a rich spectrum of scalelengths to the
hydrodynamic instabilities \index{hydrodynamic instabilities} (third snapshot in Fig.~\ref{fig1}, see also 
Fig.~\ref{fig2}a). It is not a surprise that this interaction is favoured in the presence of a large
number of bubbles. The distribution of nuclei in velocity and space for model
B30U is shown in Figs.~\ref{fig2}b and \ref{fig3}. Our results agree
qualitatively with those obtained in \cite{rhn02}.

In summary, our best model (B30U) fails to give the magnitudes adequate for a
typical SNIa. A possible cause is a poor
representation of the subsonic flame at low densities (a common problem
in most SNIa hydrocodes). However, a thorough
exploration of the parameter space of initial conditions is in order.
\noindent {\small This work has been supported by the MCYT grants EPS98--1348 
and AYA2000--1785 and by the DGES grant PB98-1183--C03-02.}

%INDEX%%%%%%%%%%%%%%%%%%%%%%%%%%%%%%%%%%%%%%%%%%%%%%%%%%%%%%%%%%%%%%%
% Please check with the editor of your book whether he plans to
% include a "mutual" subject index - if so, please code your entries
% in the standard syntax. For your own purposes you may print your
% "personal" index by using the following commands:
%
%\clearpage
%\addcontentsline{toc}{section}{Index}
%\flushbottom
%\printindex
%%%%%%%%%%%%%%%%%%%%%%%%%%%%%%%%%%%%%%%%%%%%%%%%%%%%%%%%%%%%%%%%%%%%%

\end{document}